\documentstyle[twoside,12pt,bezier,emlines,epsf]{article}
\begin{document}
\pagenumbering{Roman}
\setcounter{page}{7}

%\tableofcontents

\newpage
\pagenumbering{arabic}

\begin{center}
{\LARGE\bf                        Electronic Transport Properties                                                                  \\[2mm]
                                      of Quasicrystalline Thin Films                                                      }             \\[8mm]

{\rm                                          Roland Haberkern                                                              }               \\
{\small          Institut f\"ur Physik, TU Chemnitz, D-09107 Chemnitz, Germany         }               \\
\end{center}

%\section{Abstract}
\begin{abstract}
Quasicrystals are assumed to be electronically stabilized by a Hume-Rothery type 
mechanism. This explains most of the peculiar properties of quasicrystals. The 
stabilization is investigated by electronic transport properties, as they depend 
sensitively on the stabilizing interaction between the static structure and the 
conduction electrons.                                                                                                     
 
Thin-film techniques provide samples which are well suited for systematic 
investigations as a function of composition and structural quality for Al-Cu-Fe and 
Al-Pd-Re $i$-phases. For a narrow range of composition, large transport anomalies 
occur, reaching a metal-insulator-transition in thin films of $i$-Al-Pd-Re. We discuss 
this in the framework of a resonant scattering of the conduction electrons with the 
quasicrystalline structure, leading to a reduced electronic mobility and density of
states (DOS) at the Fermi energy $E_F$. 
\end{abstract}

\section{Introduction}
Electronic transport properties (see also \cite{May99,Lai99}) as for example the 
electrical conductivity, the 
thermopower, as well as the Hall effect are of particular interest due to their 
sensitivity to the interrelation between atomic structure and electronic system. As 
described in more detail in the contribution of H\"aussler \cite{Hau98}, a correspondence 
between the wave length of the conduction electrons at $E_F$ and a 
frequently occurring distance in the atomic system leads to resonance-like interaction, 
causing an energy-lowering of the electronic system. This behaviour can, on one 
hand, stabilize an atomic structure like the quasicrystalline one and, on the other 
hand, it modifies the electronic system so that transport anomalies arise. Stable 
quasicrystals can exhibit transport properties which can neither be described as metallic nor as 
semiconducting (activated behaviour) as shown in fig.\ref{debye}. The anomalies can be strong enough to cause a 
metal-insulator transition (MIT) in icosahedral Al-Pd-Re \cite{Pie93}. 

First investigations on quasicrystalline thin films were performed in order to 
investigate properties which could be relevant for technical applications, as for 
example mechanical, tribological and corrosion properties \cite{Mas96,Sor96}. 
Additionally, thin films were used to measure transport properties, because 
of two reasons: first of all, thin films show a much higher electrical resistance than 
bulk samples and allow a much easier and more precise determination of the 
resistivity. Films were used to determine the very small magnetoresistance at low 
magnetic fields, which is useful to interpret some transport anomalies in the framework 
of quantum corrections  \cite{Kle94}. Secondly, some work was performed to investigate 
the transition from 3 to 2-dimensional behaviour as a function of film thickness 
\cite{Yos95}.
   
We want to show here that thin quasicrystalline films do not differ substantially from bulk materials, 
but offer many advantages with respect to the preparation as well as to measuring 
electronic transport properties. Additionally, thin films can be prepared in the 
amorphous phase ($a$-phase) and crystallized afterwards into the icosahedral phase ($i$-phase) 
without the intermediate formation of crystalline counterparts. Accordingly, one 
sample allows the comparison of the quasicrystalline with the amorphous phase as 
well as the investigation of the transition between both. 

Thin-film techniques provide samples which are well suited for systematic 
investigations as a function of composition and structural quality. We show that for a 
narrow range of composition large transport anomalies occur, up to insulating films for $i$-Al-Pd-Re. 
We discuss this in the framework of resonant scattering of the 
conduction electrons with the quasicrystalline structure, causing a reduced electronic 
mobility and DOS at $E_F$.

We should address the question, which films are denoted as thin. The thickness
should be compared to the length scales of structural units or 
relevant physical lengths. The smallest structural units, namely the building clusters 
of quasicrystals are of order 1 nm \cite{Jan99}. As clusters are selfsimilar, an upper limit for 
cluster sizes is not easy to specify, but 20 nm is a value of a quite large cluster shown 
in  \cite{Jan97,Jan99}. Most of the relevant length scales of physical mechanisms are much shorter.
The electronic scattering length for example is in the order of some atomic distances in the case of 
elastic scattering (intrinsic to good quasicrystals), while the inelastic scattering length 
increases with decreasing temperature and can reach $l_i=10^2\rm nm$ at $T=1$\,K  \cite{Hab93}. 
From this it can be concluded that the thickness of a typical film is well inside the 
regime where a 3-dimensional behaviour can be expected.
A film may be called `thin' when it is prepared by a deposition process from the vapor phase.

Very thin films with a thickness starting at about $d=10$ nm were produced as 
selective solar absorbers  \cite{Eis96}. It is not clear if films of such a thickness are 
homogeneous or behave more like a percolating network. Therefore, samples for 
electronic transport measurements cover the thickness range $30<d<1000$ nm. The 
thickness of coatings for technical applications is much higher, up to 200 $\mu$m  \cite{Mas96}.

\section{Preparation of thin quasicrystalline films}
The films considered here are prepared by evaporation or sputtering from one or 
more sources onto substrates at different temperatures. 
Two routes are possible: 
homogeneous samples may be prepared directly, or crystalline multilayers of the 
different components may be prepared first, followed by extensive heat treatments. 
While the first technique can lead directly to quasicrystalline (qc) films (if deposited at a 
sufficiently high temperature), the second needs interdiffusion between the layers 
and a solid state reaction to form a qc-film. 
First films for conductivity measurements were fabricated with the second technique 
by a sequential sputtering technique of Al, Fe and Cu. As the thickness of each layer 
could be controlled by a quartz oscillator with high accuracy, the overall composition has been adjusted 
such that the narrow range of the $i$-phase in the phase diagram was reached  \cite{Kle94,Gir98}. Some groups 
have prepared thin films of Al-Cu-Fe by sputtering from a composite or area-
sectional target  \cite{Chi92,Eis96}. This results in samples with homogeneous 
composition and can lead to amorphous films if the substrate temperature is held below 
$T \approx500$\,K and to qc-films for $T>750$\,K. 
\begin{figure}[h]
 \begin{center}
 \unitlength1cm
  \begin{minipage}[t]{6cm}
  \epsfxsize=6cm
  \vspace*{0.5cm}
  \epsffile{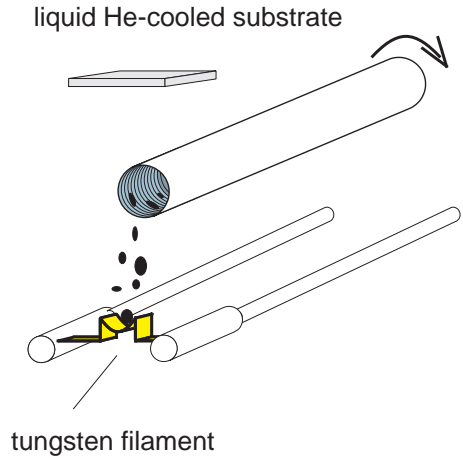}
  \end{minipage}
 \end{center}
\caption{\label{flash} Sketch of a sequential flash evaporation}
\end{figure}
The amorphous films could be transformed 
to the qc state (780 K) via a cubic phase ($T_{cryst}=580$\,K) by an appropriate heat 
treatment  \cite{Chi92}. A problem of this technique in order to gain samples with 
systematic varying compositions is the need of an individual target for each composition desired.
\begin{figure}[t]
 \begin{center}
 \unitlength1cm
  \begin{minipage}[t]{11.5cm}
  \epsfxsize=11.5cm
  \epsffile{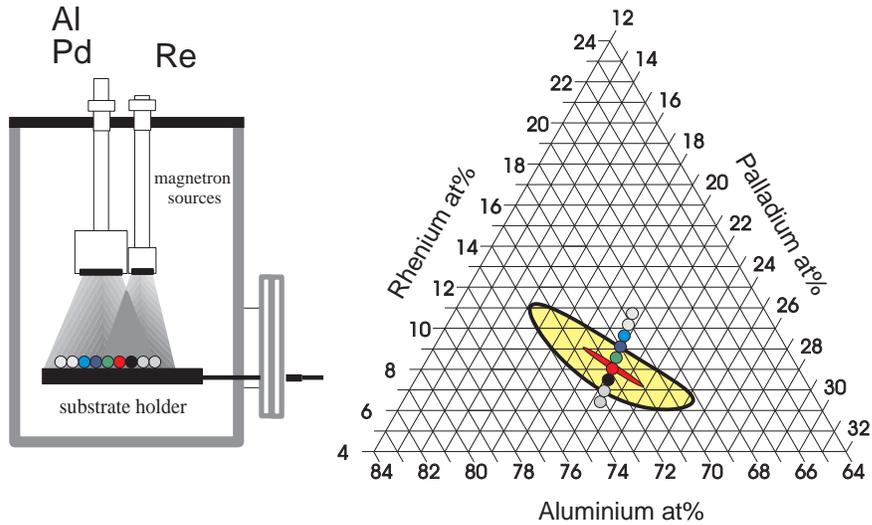}
  \end{minipage}
 \end{center}
\caption{\label{sputter2} Sketch of the co-sputtering device {\bf (a)} and a part of the
ternary phase diagram {\bf (b)} showing the 
presumed area of the $i$-phase (shadowed) and a sequence of samples
co-sputtered in one deposition step }
\end{figure}
The evaporation of a master alloy by an electron beam is another method which has 
been used  \cite{Yos95}. Here the films have compositions which are different from the 
master alloy and have to be determined afterwards. 

In the present paper we report on a sequential flash-evaporation technique (fig.\ref{flash}) in 
order to prepare amorphous Al-Cu-Fe samples with an exactly defined stochiometry 
as a precursor for the qc-phase. Grains of a premolten master alloy of the correct 
composition are feeded to a hot tungsten filament where they flash evaporate. As each 
grain contributes one monolayer or less to the film thickness, even a complete 
segregation finally ends up with a homogeneous film of the nominal composition.
As a second technique for Al-Pd-Re we use co-sputtering with two magnetron 
sources (fig.\ref{sputter2}a). Due to the positions of the two sources in respect to the substrate a defined 
composition gradient can be achieved along the substrate. With the latter technique 
in one preparation process a set of amorphous samples can be produced consisting of 
about 20 samples with a composition, slightly and systematically changing from one 
sample to the next, cutting the ternary phase diagram at, or close to the optimal 
composition (fig.\ref{sputter2}b). A defined shape of the evaporated or sputtered samples can be 
achieved by applying a mask or a structuring technique using microelectronic 
technologies, respectively. This allows to determine the absolute resistivity, which is 
difficult for most bulk samples because of their irregular shape and brittleness.

\section{Amorphous to quasicrystalline transition}
The amorphous state prepared by quenching of the vapor phase onto a cooled 
substrate is structurally similar to the liquid phase. Therefore, a direct transition from 
the amorphous to the qc-phase is possible for films at appropriate composition. In 
contrast to the transition from the liquid to the quasicrystal, the transition from the 
amorphous to the qc state is irreversible and occurs at much lower temperatures. 
\begin{figure}
 \begin{center}
  \unitlength1cm
  \begin{minipage}[t]{11.5cm}
   \epsfxsize=11.5cm
   \epsffile{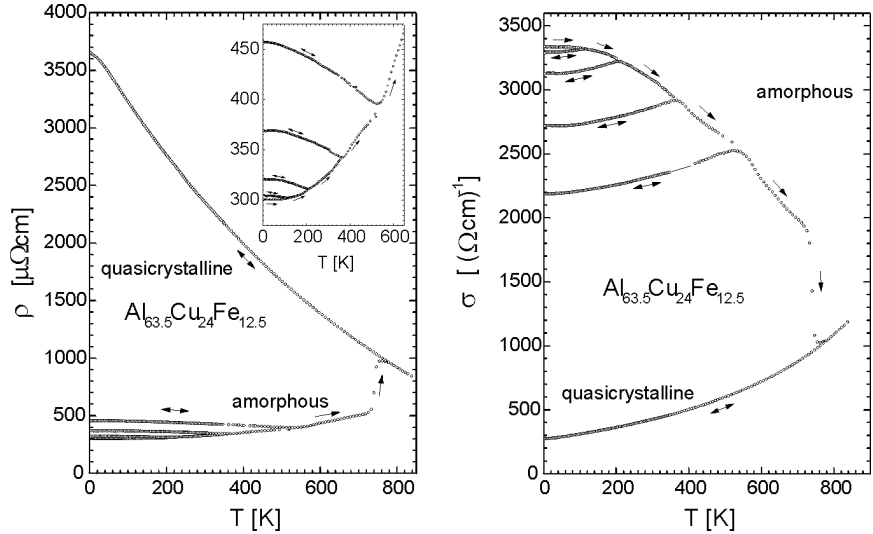}
  \end{minipage}
 \end{center}
 \caption{\label{acf_rs}Electrical resistivity {\bf(a)} and conductivity {\bf(b)} of a thin film of Al-Cu-Fe prepared at $T=4$\,K in the amorphous state. After some annealing steps in the still amorphous sample,
(enlarged in the inset) the crystallization to the 
$i$-phase occurs at $T=760$\,K  \cite{Hab98a} }
\end{figure} 
\begin{figure}[t]
 \begin{center}
 \unitlength1cm
  \begin{minipage}[t]{8cm}
  \epsfxsize=8cm
  \vspace*{0.5cm}
  \epsffile{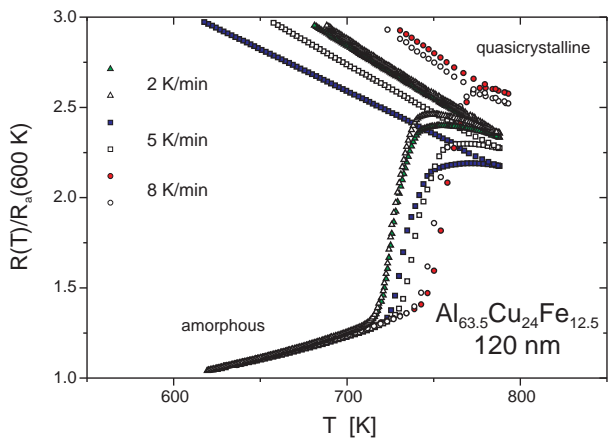}
  \end{minipage}
 \end{center}
\caption{\label{rate2}Normalized electrical resistivity of Al-Cu-Fe as a function of the heating rate for the transition 
from the amorphous to the icosahedral phase  \cite{Hab99b}}
\end{figure}   
Fig.\ref{acf_rs}a shows the resistivity of a thin film of Al-Cu-Fe deposited by sequential flash 
evaporation onto a substrate cooled down to $T=4.2$\,K. The resistivity of the amorphous 
state at this temperature is $\rho(4.2\,K)=300\,\mu\Omega$cm and increases to a value 
of $\rho(4.2\,K)=470\mu\Omega$cm after an annealing up 
to $T=550$\,K. This behaviour is discussed later. Besides the quite small and continuous 
irreversible increase of the resistivity and its temperature coefficient, a 
distinct increase related to the direct transition from the $a$- to the $i$-phase
can be seen at $T\approx740$\,K. Above, the sample is icosahedral as shown 
by electron diffraction \cite{Hab99b}. The transition temperature is about 350\,K lower than the 
solidus temperature of the same system. This offers new possibilities. First, the 
transition happens within the solid phase and allows the measurement of electronic 
transport properties at the transition (fig.\ref{rate2}).
Second, the transition plotted in fig.\ref{acf_rs} shows no indication of the 
formation of crystalline phases like the cubic $\beta$- or $\tau$-phases which occur whenever an 
Al-Cu-Fe melt is cooled down from the liquid state by melt-spinning or conventional casting. Therefore, in 
films produced via the amorphous route no long-term annealing treatments (typically some
hours at $T=1050$\,K for bulk samples) are necessary for removing crystalline phases. 
Third, the transition from the $a$- to the qc-phase happens not only at low 
$T$ but also on short time scales in the order of minutes. Fig.\ref{rate2} shows
this transition of $\rm{Al_{63.5}Cu_{24}Fe_{12.5}}$films for different heating rates. Together with a 
low surface roughness achieved by this preparation, this may have consequences for 
technical applications of qc coatings.

\section{Comparing the a-with the i-phase - the scattering approach}
The preparation of qc thin films via the route of the $a$-phase 
offers the possibility to compare directly the electronic system of the isotropic 
amorphous with the nearly isotropic quasicrystalline phase in one and the
same sample.

The conductivity of very different quasicrystals, stable ones with strong 
transport anomalies like $i$-Al-Cu-Fe, $i$-Al-Pd-Mn, or $i$-Al-Pd-Re, as well as simple ones (without $d$-states at the Fermi level)
with much smaller transport anomalies like $i$-Al-Mg-Zn have often been described by 
an inverse Matthiessen rule. That means that their $T$-dependencies of the 
conductivity $\sigma(T)$  
are claimed to be the sum of the conductivity $\sigma_0$ at $T\to0$\,K which 
depends on the particular system, its composition and structural quality, and on a 
temperature dependent increase $\Delta\sigma (T)$ which is very similar for all qc 
phases and samples 
$$\sigma(T)=\sigma_0 +\Delta\sigma(T).$$
This behaviour is inverse to that of ordinary metals, where the electrical resistivity instead of the conductivity shows a comparable behaviour. Astonishingly, fig.\ref{acf_rs}b (conductivity) shows a parallel behaviour of the curves belonging to 
the amorphous and the icosahedral phases. Accordingly, also the behaviour of the $a$-phase can be described by the inverse Matthiessen rule, suggesting a strong similarity between the amorphous and the 
quasicrystalline phases.

For quasicrystals theories exist which are based on the particular features of the quasicrystalline 
structure and its relation to the electronic system. For example a 
scattering induced electronic hopping between badly propagating electronic states 
 \cite{May93}, the localization tendency of electrons in selfsimiliar clusters  \cite{Jan97,Jan99} or a 
temperature dependent change of the band structure due to the decrease of elastic 
Bragg scattering at higher temperatures  \cite{Hab95a}. These models can explain only few aspects of 
electronic transport as for example the inverse Matthiessen rule. 
\begin{figure}[t]
 \begin{center}
 \unitlength1cm
  \begin{minipage}[t]{8cm}
  \epsfxsize=8cm
  \vspace*{0.5cm}
  \epsffile{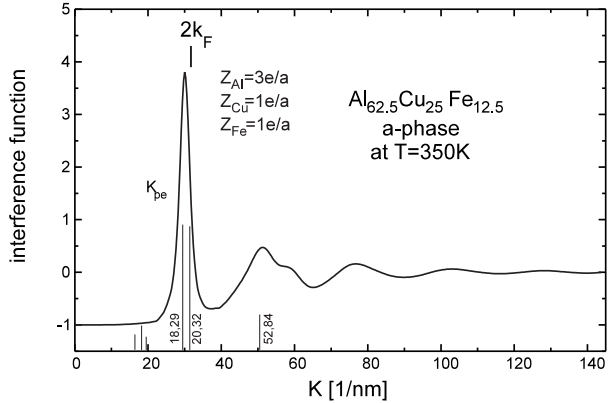}
  \end{minipage}
 \end{center}
\caption{\label{e-beug}Electron diffraction of an amorphous $\rm{Al_{62.5}Cu_{25}Fe_{12.5}}$ film (20 nm) in comparison with 
the position of major $i$-Al-Cu-Fe peaks \cite{Hab98a} }
\end{figure} 
A different approach to understand the unexpected properties of 
quasicrystals is motivated by the electronic stabilization of their peculiar structure. 
This is due to the interaction between the electrons at the Fermi level (characterized 
by $2k_F$) and the structure (characterized by the Jones zone diameter $k_{pe}$). Such a 
behaviour corresponds to the Hume-Rothery mechanism in periodic crystals  \cite{Jon37,Hum26} and was first discussed for quasicrystals by Smith and Ashcroft  \cite{Smi87} 
and Friedel  \cite{Fri88}. This description is obvious as the Jones zone of a quasicrystal 
(fig.\ref{jones}b) consists of many flat areas and hence can interact with a spherical 
Fermi surface in many directions. 
\begin{figure}[t]
 \begin{center}
 \unitlength1cm
  \begin{minipage}[t]{11.5cm}
  \epsfxsize=11.5cm
  \vspace*{0.5cm}
  \epsffile{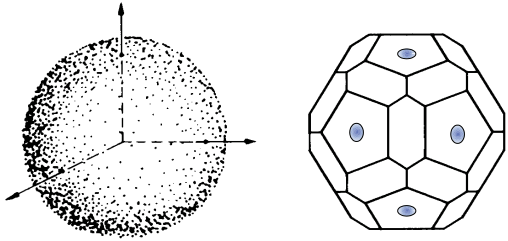}
  \end{minipage}
 \end{center}
\caption{\label{jones}Jones zone for the amorphous {\bf (a)} and the icosahedral phase {\bf (b)} constructed from the
 peak at $K_{pe}$ (fig.\ref{e-beug}) and the 20,32 and 18,29 peaks for the $a$- and $i$-phase, respectively  }
\end{figure}  
Fig.\ref{e-beug} shows that the strongest peaks of the $i$-phase are at the position of the 
electronically induced peak of the $a$-phase  \cite{Hab98a}. This suggests that the size of the Jones 
zone as well as its shape (see fig. \ref{jones}) is very similar for the $a$- and the $i$-phase. 
It has been shown  \cite{Hau94} that the stability and many other properties of a large 
class of amorphous systems can be understood by such an interaction. Thin films are 
well suited to investigate this aspect because they allow the comparison of the 
amorphous with the quasicrystalline state in one sample. Electronic transport 
properties of films can be measured in situ up to elevated temperatures due to easy preparation 
techniques and the high contact stability.

The interaction between electronic and atomic system is based on the elastic 
scattering of electrons with the Fermi wavenumber ($k=k_F$) at a pronounced peak in the 
structure factor $S(2k_F)$ describing the atomic structure. The elastic (Bragg-like) 
scattering of electrons reduces not only the mean free path, the main effect may be 
the depression of the DOS at $E_F$. If the matching of 
the $k$-vectors of the electronic and the atomic system is good, this results in a 
resonant-like scattering of electrons. The previously freely propagating electrons change 
partially to standing waves because of resonant scattering, resulting in a tendency of 
localization of the electrons at $E_F$. As a consequence, the electronic 
DOS at $E_F$ is reduced, the stability of the system increases 
and electronic transport anomalies arise. The aspect of stability is comprehensively 
discussed in  \cite{Hau98}, while the consequences for electronic transport in amorphous 
and quasicrystalline films will be briefly discussed in the following.
In the case of a weak scattering approach the Ziman formula 
$$\rho_z={1\over\sigma_z}\propto\int_0^{2k_F}S(K)\vert v(K)\vert ^2 K^3 dk$$
 \cite{Zim70} has been used to describe the resistivity and its temperature 
dependence for liquid and amorphous systems. $K$ represents the scattering vector,
 $v(K)$ the pseudopotential and $S(K)$ 
the static structure factor. This formula cannot strictly be applied to 
systems with strong scattering as are some amorphous systems or 
quasicrystals. But this relatively simple formula can be helpful to follow up the 
trends to stronger scattering. Due to the weighting with $K^3$ the elastic 
scattering strongly depends on $S(2 k_F)$, the structural weight at the upper limit of the 
integral. Thus, a pronounced peak in the structure factor at $2 k_F$ leads to a strong 
elastic scattering and a low conductivity. As the temperature increases, 
the portion of the elastic scattering at $K=2 k_F$ is reduced in favour of inelastic 
scattering. The latter is less effective on reducing the conductivity than the 
diffuse umklapp scattering. Thus, 
the conductivity  increases with increasing inelastic scattering due to rising temperature.
\begin{figure}[t]
 \begin{center}
 \unitlength1cm
  \begin{minipage}[t]{11.5cm}
  \epsfxsize=11.5cm
  \vspace*{0.5cm}
  \epsffile{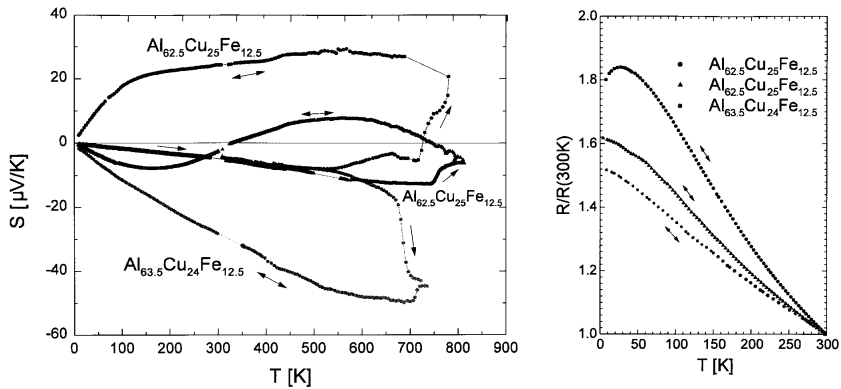}
  \end{minipage}
 \end{center}
\caption{\label{acf_ther}Thermopower {\bf (a)} and normalized resistance {\bf(b)} for 3 Al-Cu-Fe films of slightly different composition  \cite{Rot99}}
\end{figure}  
Additionally, the resonant like elastic scattering due to the coincidence of the 
electronic wavelength with structural distances opens a gap or a pseudogap at $E_F$. 
This is similar to the case of periodic crystals where a matching of the wave length of 
the electron waves at $E_F$ with the wave length of a reciprocal lattice vector causes a 
gap due to the energy difference for the two solutions of standing electron waves. In 
quasicrystals the matching may be slightly different. A resonance-like interaction is  
most effective, if not only the wavelength but also the waveforms of both systems are 
identical. In quasicrystals and especially in amorphous alloys, it is well known that 
the atoms are onionshell-like located around ad-atoms. The distance between 
neighbouring shells $\lambda _P$ is in agreement with the Friedel wavelength 
$\lambda _F=2\pi/2k_F$. Each 
atom is surrounded by `mirror spheres' analog to the equidistant separation of mirror 
planes in the crystalline case. The electrons should be treated as spherical waves in 
coincidence with the onionshell-like static structure. Under resonance, any plane 
wave which once may have been formed will by scattering immediately fall apart 
into spherical waves which are strongly localized due to the phase-coherent 
backscattering from all the neighbouring onionshells. The resonance-like coupling 
causes localization of the electrons at any site and hence a short mean-free path 
which itself helps to enhance the role of the spherical waves at medium range. In the 
electronic DOS pseudogaps arise. The onionshell-like position of the neighbouring 
atoms around any ad-atom may also be seen as a cluster-model with strongly 
interpenetrating clusters.
Whereas in the amorphous phase the location of the atoms on the onionshells is still 
strongly random and in addition angular correlations are weak or absent, this is no 
longer the case in the qc-phase. The distances of the individual atoms to the ad-atoms are 
much better defined and strong angular correlations exist. The phase coherence of 
the backscattered spherical waves is better fulfilled and therefore the interference 
effects are even stronger. Accordingly the localization of the electron waves should 
be stronger and the pseudogap should become deep. 
As the temperature rises the portion of the resonant elastic scattering is decreased in 
favour of inelastic scattering reducing the pseudogap both for amorphous and 
quasicrystalline systems. We mention that this behaviour is simply ruled by the 
Debye-Waller factor which describes the ratio of the elastic to inelastic scattering for 
the relevant peaks building up the Jones zone. This can be seen in fig.\ref{acf_rs}b. The 
conductivity for all reversible parts (marked by $\leftrightarrow$ ) of the amorphous states and also for the 
quasicrystalline state increases with temperature in exactly the same manner. The 
conductivity irreversibly decreases (marked by $\searrow$) with every annealing step reaching its smallest value in the 
qc-phase with its strong radial and angular correlation of atoms. The temperature 
dependence of the conductivity is discussed in detail in the Al-Pd-Re system as this 
icosahedral system shows even larger transport anomalies (reaching insulating behaviour) and 
has a much smaller contribution of low temperature effects from spin-orbit scattering. 

The effect of resonance in amorphous and quasicrystalline samples can also be seen 
in the thermopower. For amorphous systems it is well known that a stronger 
resonance between static structure and electronic system causes a deviation from the 
free electron value of the thermopower $S_0$ to positive values  \cite{Hau94}. Fig.\ref{acf_ther} shows the same behaviour 
for the icosahedral phase. The samples show a transition from a large negative to a 
large positive thermopower. This correlates with the increase of the resistance ratio 
$\rho(4K)/\rho(300K)$, which is an indication of the enhanced resonance effect between 
static structure and electronic system. Note that the $i$-phases show thermopowers 
which are much larger than the free-electron value. This can be attributed to a deep 
pseudogap which reduces the effective number of charge carriers at the Fermi-level.

\section{The Al-Pd-Re system}
The icosahedral Al-Pd-Re system shows the largest transport anomalies amongst all 
the quasicrystals, up to a metal-insulator transition (MIT)  \cite{Pie93}. Thus it is interesting 
to check the concept of resonant interaction between the static 
structure and the conduction electrons on this alloy. Due to the extremely different vapor 
pressures of the constituents it is difficult to produce homogeneous bulk samples of a 
defined composition in a reproducible way. With the co-sputtering technique 
mentioned before, systematic and slightly different samples can be fabricated as thin films. 
\begin{figure}[h]
 \begin{center}
 \unitlength1cm
  \begin{minipage}[t]{11cm}
  \epsfxsize=11cm
   \epsffile{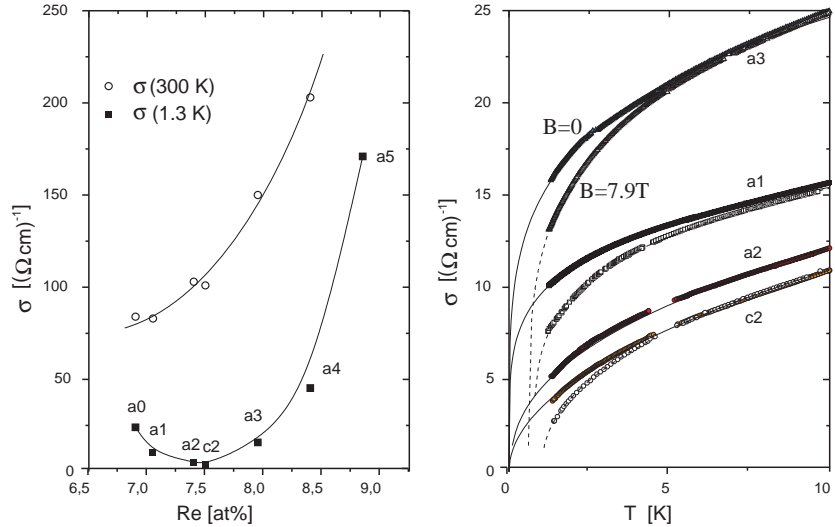}
  \end{minipage}
 \end{center}
\caption{\label{sig_re}Conductivity for a sequence of Al-Pd-Re films of constant Al/Pd
ratio {\bf(a)} and their low temperature 
conductivity {\bf(b)} with a magnetic field of $B=7.9$T (open symbols) and without field (solid symbols)}
\end{figure}
As an example, Fig.\ref{sig_re}a shows the conductivity of such a sequence as a function of the Rhenium 
content for a constant Aluminium to Palladium ratio. There exists a systematic 
variation of the low $T$ conductivity with a minimum around $\rm{Al_{72.3}Pd_{20.2}Re_{7.5}}$. 
Simultaneously, some of the samples seem to be insulating  \cite{Hab99}, that means, the 
conductivity vanishes for $T\to0$\,K (fig. \ref{sig_re}b). A magnetic field enhances this behaviour 
and apparently leads to a vanishing conductivity at a small but finite temperature.

\begin{figure}[h]
 \begin{center}
 \unitlength1cm
  \begin{minipage}[t]{11.5cm}
  \epsfxsize=11.5cm
  \vspace*{0.5cm}
  \epsffile{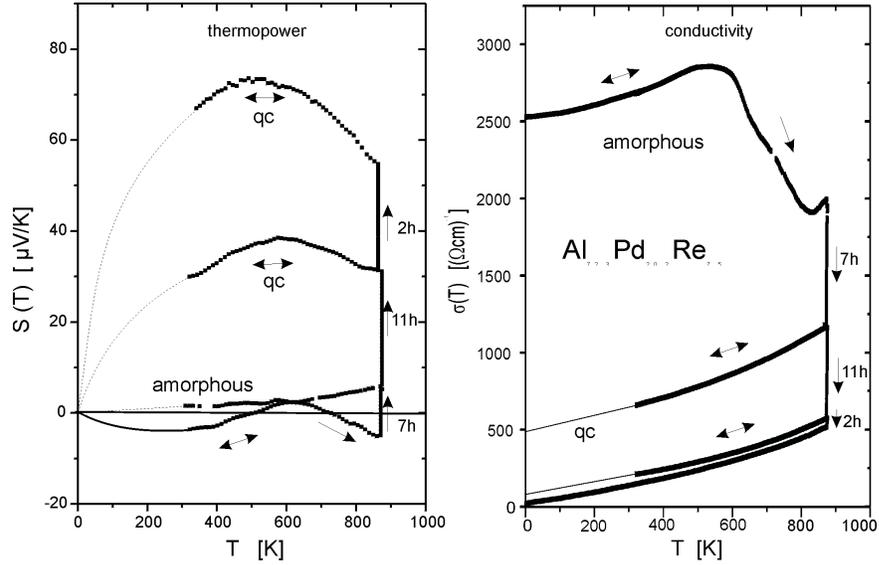}
  \end{minipage}
 \end{center}
\caption{\label{ssr1677}Thermopower {\bf (a)} and conductivity {\bf(b)}
for a Al-Pd-Re film during annealing and in steps of different 
structural quality }
\end{figure}

Thereby, the transport properties of $i$-Al-Pd-Re match into the common behaviour of stable quasicrystals mentioned before in the $i$-Al-Cu-Fe system. Fig.\ref{ssr1677}b shows again the validity of the inverse Matthiessen-rule for the conductivity of the $a$- and $i$-phases of different structural qualities. In contrast to all other qc systems, the Al-Pd-Re system offers two exciting aspects.  First, the conductivity vanishes at $T\to0$\,K, which may be due a stronger resonant 
scattering because of heavier elements in the case of Al-Pd-Re. Second, at low $T$, there is no increase of conductivity, as for high quality $i$-Al-Cu-Fe (fig.\ref{acf_ther}b) or for other stable quasicrystals like $i$-Al-Pd-Mn. Such a low $T$ increase of conductivity (attributed to strong spin-orbit scattering and electron-electron interaction  \cite{Hab93}) makes a MIT unlikely and complicates interpretation.
\begin{figure}[h]
 \begin{center}
 \unitlength1cm
  \begin{minipage}[t]{11.5cm}
  \epsfxsize=11.5cm
  \vspace*{0.5cm}
  \epsffile{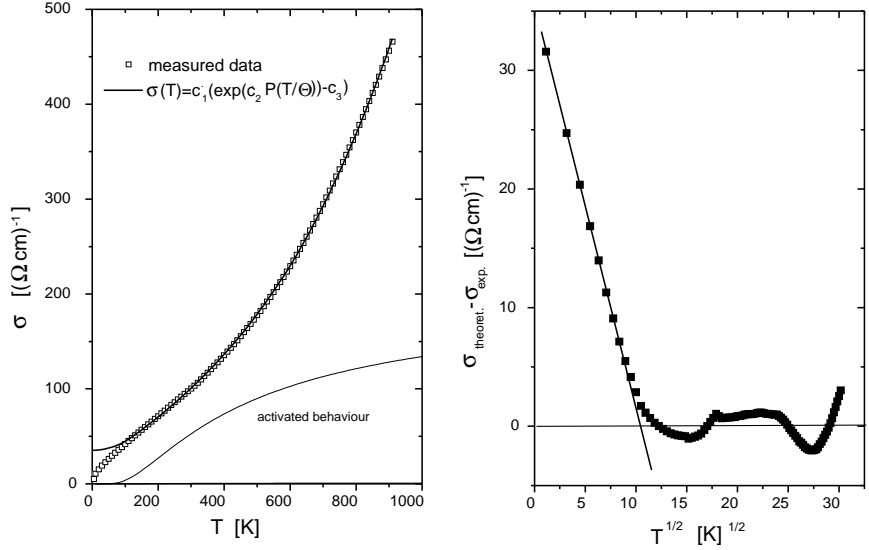}
  \end{minipage}
 \end{center}
\caption{\label{debye}Conductivity of a $\rm{Al_{72.3}Pd_{20.2}Re_{7.5}}$ film 
in comparison with a fit according to a function of the 
Debye-Waller factor {\bf(a)} and difference fit and experimental values {\bf(b)} }
\end{figure}  
The model of a resonant-like interaction of the electronic and the static structure is able 
to explain the transport properties also for the insulating samples. Fig.\ref{debye} shows a fit to 
the conductivity data in a quite large temperature range. The fit is indirectly
proportional to a function of the Debye-Waller factor $exp(-cP({T\over \Theta}))$, with the
Debeye-temperature $\Theta$ and a constant $c$ which describes the 
temperature dependent decrease of the elastic scattering of the electrons at $2k_F$ at the 
Jones zone. There has been only one parameter ($c_3$\,in the expression written in
fig.\ref{debye}) added which is related to the size of 
the gap between electron- and hole-like charge carriers and is related to the effective 
pseudopotential. A significant difference between fit and measured data occurs only 
at temperatures below 100 K. Fig.\ref{debye}b shows that this difference is proportional to $\sqrt T$ 
which can be attributed to electron-electron interaction (EEI). In systems of strong 
elastic scattering, electrons are less able to screen charge variations and EEI is 
enhanced, leading to a small dip in the electronic density of states at $E_F$ \cite{Alt79}. In the case 
of amorphous alloys this behaviour can be computed quantitatively as a function of 
temperature and magnetic field  \cite{Alt83,Lee82}. In weakly insulating icosahedral 
systems the MIT seems to have two origins. First, the resonance-like interaction 
between the static structure leads to a deep pseudogap in the electronic DOS at $E_F$ 
and the localization tendency of the conduction electrons due to standing spherical 
waves in some structural clusters. Second, the EEI lowers the small DOS at $E_F$ 
additionally. The sum of both effects leads to a vanishing conductivity at $T\to 0$\,K. A magnetic field dephases the electronic wavefunction enlarging the EEI by a further decreased screening. The MIT occurs at a finite temperature.

The thermopower of $i$-Al-Pd-Re can be described in the resonance-model mentioned 
before. As shown in fig.\ref{ssr1677}, the thermopower increases to large positive values as the 
structural quality increases from the amorphous to the defective and finally the good 
icosahedral state. The conductivity of the same sample reduces in the same manner 
as the resonance effect increases.

\section {Conclusion}
Quasicrystalline thin films can be produced with high structural quality. The low 
temperature electrical conductivity, which is very sensitive to the structure quality, 
reaches values which are comparable to that of bulk materials. A MIT is achieved in 
both, the bulk samples as well as in thin films of $i$-Al-Pd-Re. Thin films provide the 
possibility to make systematic investigations as a function of composition and 
structural quality. 
Further understanding in the interpretation of the peculiar transport anomalies of 
quasicrystals could be achieved by the comparison to the isotropic amorphous phase. 
This has been done on different annealing states of the same sample. Astonishingly, 
the temperature dependence of the conductivity is qualitatively very similiar for both 
phases, the inverse Matthiessen rule which is thought to be peculiar to quasicrystals 
is also valid for amorphous samples of the same composition. We discussed this in 
the framework of a resonant scattering between the conduction electrons and the 
static structure. Hereby, the mechanism is the same for amorphous as well as for qc 
systems. Due to the sharp distance distributions and angular correlations in 
quasicrystals the quantitiy of the scattering is much larger in the quasicrystalline 
case, leading to a deeper pseudogap and a stronger localization tendency of the 
conduction electrons. The temperature dependence of the conductivity could be 
fitted due to the scattering approach by a simple function of the Debye-Waller factor 
for the temperature range from 100 to 900\,K. At low temperatures the conductivity 
of $i$-Al-Pd-Re is additionally influenced by strong electron-electron interaction. This 
leads to an additional dip of the electronical DOS at $E_F$ and finally to the metal-
insulator-transition. 
The thermopower increases to large positive values as the conductivity vanishes for 
high-quality qc samples. Again, this is a typical behaviour of electronically stabilized 
systems, which can be described by a strong interaction between the conduction 
electrons and the static structure. 

\section{Acknowledgements}
Support of this work by DFG under contracts Ha2359/1, Ha1627/8 and Ha1627/9 is gratefully acknowledged.

%\newpage
%\addcontentsline{toc}{chapter}{Index}
%\thispagecropped

%\printindex

\begin{thebibliography}{99}
%\addcontentsline{toc}{section}{References}

\bibitem{Alt79}    Altshuler B.L., Aronov A.G., 
                               {\it Solid State Comm.} {\bf 30} (1979) 115
\bibitem{Alt83}    Altshuler B.L., Aronov A.G., 
                               {\it Solid State Comm.} {\bf 46} (1983) 429
\bibitem{Chi92}	Chien C.L., Lu M.,
                               {\it Phys. Rev. B} {\bf 45} (1992) 12793
\bibitem{Eis96}	Eisenhammer T., Mahr A., Haugeneder A., Reichelt T., Assmann W., 
	                    Proceeddings of the 5th Int. Conf. on Quasicrystals. Eds. C. Janot, R. 
                                    Mosseri, World Scientific (1996) 758
\bibitem{Fri88}	Friedel J.,  
                                {\it Helv. Phys. Acta} {\bf 61} (1988) 538
\bibitem{Gir98}	Giroud F., Grenet T., Schaub T.M., Berger C., Barna B.P., Radi Z., 
                                Proceeddings of the 6th Int.Conf. on Quasicrystals. Eds. Takeuchi S., Fujiwara T., 
                                World Scientific (1998) 712
\bibitem{Hab93}	Haberkern R., Fritsch G., Schilling J.,  
                               {\it Z. Phys. B} {\bf 92} (1993) 383
\bibitem{Hab95a} Haberkern R., Fritsch G., 
                                Proceeddings of the 5th Int.Conf. on Quasicrystals. Eds. Janot C., Mosseri R., World Scientific (1996) 460
\bibitem{Hab98a}	 Haberkern R., Roth C., Kn\"ofler R., Zavaliche F., H\"aussler P., 
                                 Proceedings of the 6th Int.Conf. on Quasicrystals. Eds. S. Tekeuchi, T. Fujiwara, World Scientific (1998) 643
\bibitem{Hab98b}	 Haberkern R., Roth C., Kn\"ofler R., Zavaliche F., H\"aussler P., 
                                 Proceedings of the 6th Int.Conf. on Quasicrystals. Eds. Tekeuchi S., Fujiwara T., World Scientific (1998) 716
\bibitem{Hab99b}	Haberkern R., Roth C., Kn\"ofler R., Schulze S., H\"aussler P.,
                                in: Quasicrystals, MRS-Proceedings Vol. 553 (1999), eds. Dubois J.-M., Thiel P. A., Tsai P.-A., Urban K.  
\bibitem{Hab99}	Haberkern R., Rosenbaum R., H\"aussler P.,
                                to be published
\bibitem{Hau94}	H\"aussler P.,
                                in: Glassy Metals III, eds Beck H., G\"unterodt H.-J. Topics in Applied Physics (Springer), Vol. 72 (1994) 163
\bibitem{Hau98}  H\"aussler P., 
                               "Quasicrystals - material on the path from disorder to order", this volume
\bibitem{Hum26} Hume-Rothery W.,  
                               {\it J. Inst. Met.}  {\bf35} (1926) 295	
\bibitem{Jan97}	Janot C., J. Phys.,
                               {\it Condens. Matter} {\bf 9} (1997) 1493
\bibitem{Jan99}	Janot C., Dubois J.-M., 
                                "Quasicrystals as hierarchical packing of overlapping clusters", this volume
\bibitem{Jon37}	Jones H., 
                               {\it Proc. Phys. Soc.}  {\bf49} (1937) 250
\bibitem{Kle94}	Klein T, Symko O.G., 
                               {\it Appl Phys. Lett.} {\bf 64} (1994) 431
\bibitem{Lee82}	Lee P.A., Ramakrishnan T.V.,  
                               {\it Phys. Rev. B}  {\bf26} (1982) 4009
\bibitem{Mas96}	Massiani Y., Ait Yaazza S., Dubois J.-M., 
                                Proceedings of the 5th Int. Conf. on Quasicrystals. Eds.  Janot C., Mosseri R., World Scientific (1996) 790
\bibitem{May93}	Mayou D., Berger C., Cyrot-Lackmann F., Klein T., Lanco P.,  
                                {\it Phys. Rev. Lett.}{\bf 70} (1993) 3915
\bibitem{Pie93}	Pierce F.S., Poon S.J., Guo Q.,  
                               {\it Science} {\bf 261} (1993) 737
\bibitem{May99}	Roche S., Mayou D., 
                                "Electronic conductivity of quasicrystals and approximants", this volume
\bibitem{Rod96}	Rodmar M., Ahlgren M., Rapp \"O., 
                                Proceedings of the 5th Int. Conf. on Quasicrystals. Eds. Janot C., Mosseri R., World Scientific (1996)  518
\bibitem{Rot99}	Roth C., Schwalbe G., Kn\"ofler, Zavaliche F., Madel O., Haberkern R., H\"aussler P., 
                                Proceedings of the 10th Int. Conf. on Liquid and Amophous Metals (1998), to be published 	 
\bibitem{Smi87}	Smith A.P., Ashcroft N.W.,  
                                {\it Phys. Rev. Lett.} {\bf 59} (1987) 1365
\bibitem{Sor96}	Sordelet D.J., Kramer M.J., Anderson I.E., Besser M.F., 
                                Proceedings of the 5th Int. Conf. on Quasicrystals. Eds. Janot C., Mosseri R., World Scientific (1996) 778
\bibitem{Lai99}	Trambly de Laissardi\`ere G., Mayou D., 
                                "Magnetic properties of quasicrystals and approximants", this volume 
\bibitem{Yos95}	Yoshioka A., Edagawa K., Kimura K., Takeuchi S.,  
                                {\it Jpn. J. Appl. Phys.}  {\bf34} (1995) 1606
\bibitem{Zim70}	Ziman J.M., 
                                {\it  Proc. Roy. Soc. , Ser. A} {\bf 318} (1970) 401
\end{thebibliography}
\end{document}